# Lessons learned from the deployment of a high-interaction honeypot

E. Alata[1], V. Nicomette[1], M. Kaâniche[1], M. Dacier[2], M. Herrb[1]

[1]LAAS-CNRS, University of Toulouse, 7 Avenue du Colonel Roche, 31077 Toulouse Cedex 4, France
[2]Eurécom, 2229 Route des Crêtes, BP 193, 06904 Sophia Antipolis Cedex, France
{alata, nicomette, kaaniche, herrb}@laas.fr; dacier@eurecom.fr

**Abstract**

*This paper presents an experimental study and the lessons learned from the observation of the attackers when logged on a compromised machine. The results are based on a six months period during which a controlled experiment has been run with a high interaction honeypot. We correlate our findings with those obtained with a worldwide distributed system of low-interaction honeypots.*

## 1. Introduction

During the last decade, several initiatives have been developed to monitor and collect real world data about malicious activities on the Internet, e.g., the Internet Motion Sensor project [1], CAIDA [2] and Dshield [3]. The CADHo project [4] in which we are involved is complementary to these initiatives and is aimed at:
- deploying a distributed platform of honeypots [5] that gathers data suitable to analyze the attack processes targeting a large number of machines on the Internet;
- validating the usefulness of this platform by carrying out various analyses, based on the collected data, to characterize the observed attacks and model their impact on security.

A honeypot is a machine connected to a network but that no one is supposed to use. If a connection occurs, it must be, at best an accidental error or, more likely, an attempt to attack the machine.

The first stage of the project focused on the deployment of a data collection environment (called Leurré.com [6]) based on low-interaction honeypots. As of today, around 40 honeypot platforms have been deployed at various sites from academia and industry in almost 30 different countries over the five continents. Several analyses and interesting conclusions have been derived based on the collected data as detailed e.g., in [4,5,7-9]. Nevertheless, with such honeypots, hackers can only scan ports and send requests to fake servers without ever succeeding in taking control over them. The second stage of our project is aimed at setting up and deploying high-interaction honeypots to allow us to analyze and model the behavior of malicious attackers once they have managed to compromise and get access to a new host, under strict control and monitoring. We are mainly interested in observing the progress of real attack processes and the activities carried out by the attackers in a controlled environment.

In this paper, we describe the lessons learned from the development and deployment of such a honeypot. The main contributions are threefold. First, we do confirm the findings discussed in [9] showing that different sets of compromised machines are used to carry out the various stages of planned attacks. Second, we do outline the fact that, despite this apparent sophistication, the actors behind those actions do not seem to be extremely skillful, to say the least. Last, the geographical location of the machines involved in the last step of the attacks and the link with some phishing activities shed a geopolitical and socio-economical light on the results of our analysis.

The paper is organized as follows. Section 2 presents the architecture of our high-interaction honeypot and the design rationales for our solution. The lessons learned from the attacks observed over a period of almost 4.5 months are discussed in Section 3. Finally, Section 4 concludes and discusses future work. An extended version of this paper detailing the context of this work and the related state-of-the art is available in [10].

## 2. Architecture of our honeypot

In our implementation, we decided to use VMware [11] and to install virtual operating system upon it. Compared to solutions based on physical machines, virtual honeypots provide a cost effective and flexible solution that is well suited for running experiments to observe attacks.

The objective of our experiment is to analyze the behavior of the attackers who succeed in breaking into a machine. The vulnerability that they exploit is not as crucial as the activity they carry out once they have broken into the host. That's why we chose to use a simple vulnerability: weak passwords for `ssh` user accounts. Our honeypot is not particularly hardened for two reasons. First, we are interested in analyzing the behavior of the attackers even when they exploit a buffer overflow and become root. So, if we use some kernel patch such as *Pax* [12], our system will be more secure but it will be impossible to observe some behavior. Secondly, if the system is too hardened, the intruders may suspect something abnormal and then give up.

In our setup, only `ssh` connections to the virtual host are authorized so that the attacker can exploit this vulnerability. A firewall blocks all connection attempts from the Internet, but those to port 22 (`ssh`). Also, any connection from the virtual host to the outside is blocked to avoid that intruders attack remote machines from the honeypot. This does not prevent the intruder from downloading code, using the `ssh` connection[1].

Our honeypot is a standard Gnu/Linux installation, with kernel 2.6, with the usual binary tools. No additional software was installed except the `http apache` server. This kernel was modified as explained in the next subsection. The real host executing VMware uses the same Gnu/Linux distribution and is isolated from outside.

In order to log what the intruders do on the honeypot, we modified some drivers functions (`tty_read` and `tty_write`), as well as the `exec` system call in the Linux kernel. The modifications of `tty_read` and `tty_write` enable us to intercept the activity on all the terminals of the system. The modification of the `exec` system call enables us to record the system calls used by the intruder. These functions are modified in such a way that the captured information is logged directly into a buffer of the kernel memory of the honeypot itself.

Moreover, in order to record all the logins and passwords tried by the attackers to break into the honeypot we added a new system call into the kernel of the virtual operating system and we modified the source code of the `ssh` server so that it uses this new system call. The logins and passwords are logged in the kernel memory, in the same buffer as the information related to the commands used by the attackers.

The activities of the intruder logged by the honeypot are preprocessed and then stored into an SQL database. The raw data are automatically processed to extract relevant information for further analyses, mainly: i) the IP address of the attacking machine, ii) the login and the password tested, iii) the date of the connection, iv) the terminal associated (`tty`) to each connection, and v) each command used by the attacker.

## 3. Experimental results

This section presents the results of our experiments. First, we give global statistics in order to give an overview of the activities observed on the honeypot, then we characterize the various intrusion processes. Finally, we analyze in detail the behavior of the attackers once they break into the honeypot. In this paper, an *intrusion* corresponds to the activities carried out by an intruder who has succeeded to break into the system.

### 3.1. Global statistics

The high-interaction honeypot has been deployed on the Internet and has been running for 131 days during which 480 IP addresses have tried to contact its `ssh` port. It is worth comparing this value to the amount of hits observed against port 22, considering all the other low-interaction honeypot platforms we have deployed in the rest of the world (40 platforms). In the average, each platform has received hits on port 22 from around approximately 100 different IPs during the same period of time. Only four platforms have been contacted by more than 300 different IP addresses on that port and only one was hit by more visitors than our high interaction honeypot. Even better, the low-interaction platform maintained in the same subnet as the high-interaction honeypot experimented only 298 visits, i.e. less than two thirds of what the high-interaction did see. This very simple and first observation confirms the fact already described in [9] that some attacks are driven by the fact that attackers know in advance, thanks to scans done by other machines, where potentially vulnerable services are running. The existence of such a service on a machine will trigger more attacks against it. This is what we observe here: the low interaction machines do not have the `ssh` service open, as opposed to the high interaction one, and, therefore get less attacked than the one where some target has been identified.

The number of `ssh` connection attempts to the honeypot we have recorded is 248717 (we do not consider here the scans on the `ssh` port). This represents about 1900 connection attempts a day. Among these 248717 connection attempts, only 344 were successful. Table 1 represents the user accounts that were mostly tried (the top ten) as well as the number of different passwords that have been tested by the attackers. It is noteworthy that many user accounts corresponding to usual first names have also regularly been tested on our honeypot. The total number of accounts tested is 41530.

| Account | Number of connection attempts | Percentage of connection attempts | Number of passwords tested |
|---|---|---|---|
| root | 34251 | 13.77% | 12027 |
| admin | 4007 | 1.61% | 1425 |
| test | 3109 | 1.25% | 561 |
| user | 1247 | 0.50% | 267 |
| guest | 1128 | 0.45% | 201 |
| info | 886 | 0.36% | 203 |
| mysql | 870 | 0.35% | 211 |
| oracle | 857 | 0.34% | 226 |
| postgres | 834 | 0.33% | 194 |
| webmaster | 728 | 0.29% | 170 |

**Table 1: ssh connection attempts and number of passwords tested**

Before the real beginning of the experiment (approximately one and a half month), we had deployed a machine with a `ssh` server correctly configured, offering no weak account and password. We have taken advantage of this observation period to determine which accounts were mostly tried by automated scripts. Using this acquired knowledge, we have created 17 user accounts and we have started looking for successful intrusions. Some of the created accounts were among the most attacked ones and others not. As we already explained in the paper, we have deliberately created user accounts with weak passwords (except for the `root` account). Then, we have measured the time between the creation of the account and the first successful connection to this account, then the duration between the first successful connection and the first real intrusion (as explained in section 3.2, the first successful connection is very seldom a real intrusion but rather an automatic script which tests passwords).

---
[1] We have sometimes authorized `http` connections for a short time, by checking that the attackers were not trying to attack other remote hosts.

Table 2 summarizes these durations (`UAi` means `User Account i`).

| User Account | Duration between creation and first successful connection | Duration between first successful connection and first intrusion |
|---|---|---|
| UA1 | 1 day | 4 days |
| UA2 | Half a day | 4 minutes |
| UA3 | 15 days | 1 day |
| UA4 | 5 days | 10 days |
| UA5 | 5 days | null |
| UA6 | 1 day | 4 days |
| UA7 | 5 days | 8 days |
| UA8 | 1 day | 9 days |
| UA9 | 1 day | 12 days |
| UA10 | 3 days | 2 minutes |
| UA11 | 7 days | 4 days |
| UA12 | 1 day | 8 days |
| UA13 | 5 days | 17 days |
| UA14 | 5 days | 13 days |
| UA15 | 9 days | 7 days |
| UA16 | 1 day | 14 days |
| UA17 | 1 day | 12 days |

**Table 2: History of breaking accounts**

The second column indicates that there is usually a gap of several days between the time when a weak password is found and the time when someone logs into the system with this password to issue some commands on the now compromised host. This is a somehow a surprising fact and is described with some more details here below. The particular case of the `UA5` account is explained as follows: an intruder succeeded in breaking the `UA4` account. This intruder looked at the contents of the `/etc/passwd` file in order to see the list of user accounts for this machine. He immediately decided to try to break the `UA5` account and he was successful. Thus, for this account, the first successful connection is also the first intrusion.

### 3.2. Intrusion process

In the section, we present the conclusions of our analyses regarding the process to exploit the weak password vulnerability of our honeypot. The observed attack activities can be grouped into three main categories: 1) dictionary attacks, 2) interactive intrusions, 3) other activities such as scanning, etc.

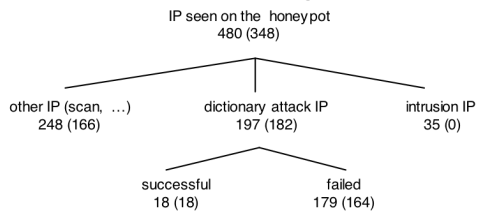

X (Y): Y IP addresses among X were also seen on the low-interaction honeypots

**Figure 3: Classification of observed IP addresses**

As illustrated in figure 3, among the 480 IP addresses that were seen on the honeypot, 197 performed dictionary attacks and 35 performed real intrusions on the honeypot (see below for details). The 248 IP addresses left were used for scanning activity or activity that we did not clearly identified. Among the 197 IP addresses that made dictionary attacks, 18 succeeded in finding passwords. The others (179) did not find the passwords either because their dictionary did not include the accounts we created or because the corresponding weak password had already been changed by a previous intruder. We have also represented in Figure 3 the corresponding number of IP addresses that were also seen on the low-interaction honeypot deployed in the context of the project in the same network (between brackets). Whereas most of the IP addresses seen on the high interaction honeypot are also observed on the low interaction honeypot, none of the 35 IPs used to really log into our machine to launch commands have ever been observed on any of the low interaction honeypots that we do control in the whole world! This striking result is discussed hereafter.

**3.2.1. Dictionary attack.** The preliminary step of the intrusion consists in dictionary attacks[2]. In general, it takes only a couple of days for newly created accounts to be compromised. As shown in Figure 3, these attacks have been launched from 197 IP addresses. By analysing more precisely the duration between the different `ssh` connection attempts from the same attacking machine, we can say that these dictionary attacks are executed by automatic scripts. As a matter of fact, we have noted that these attacking machines try several hundreds, even several thousands of accounts in a very short time.

We have made then further analyses regarding the machines that succeed in finding passwords, i.e., the 18 IP addresses. By searching the leurré.com database containing information about the activities of these addresses against the other low interaction honeypots we found four important elements of information. First, we note that none of our low interaction honeypot has an ssh server running, none of them replies to requests sent to port 22. These machines are thus scanning machines without any prior knowledge on their open ports. Second, we found evidences that these IPs were scanning in a simple sequential way all addresses to be found in a block of addresses. Moreover, the comparison of the fingerprints left on our low interaction honeypots highlights the fact that these machines are running tools behaving the same way, not to say the same tool. Third, these machines are only interested in port 22, they have never been seen connecting to other ports. Fourth, there is no apparent correlation as far as their geographical location is concerned: they are located all over the world.

In other words, it comes from this analysis that these IPs are used to run a well known program. The detailed analysis of this specific tool is outside the scope of the paper but, nevertheless, it is worth mentioning that the activities linked to that tool, as observed in our Leurré.com database, indicate that it is unlikely to be a worm but rather an easy to use and widely spread tool.

**3.2.2. Interactive attack: intrusion.** The second step of the attack consists in the real intrusion. We have noted that, several days after the guessing of a weak password, an interactive `ssh` connection is executed on our honeypot to issue several commands. We believe that, in those situations, a real human being, as opposed to an automated script, is connected to our machine. This is explained and justified in Section 4.3. As shown in Figure 3, these intrusions come from 35 IP addresses never observed on any of the low-interaction honeypots.

---
[2] We consider as "dictionary attack" any attack that tries more than 10 different accounts and passwords.

Whereas the geographic localisation of the machines performing dictionary attacks is very blur, the machines that are used by a human being for the interactive `ssh` connection are, most of the time, clearly identified. We have a precise idea of their country, geographic address, the responsible of the corresponding domain. Surprisingly, these machines, for half of them, come from the same country, an European country not usually seen as one of the most attacking ones as reported, for instance, by the www.leurrecom.org web site.

We then made analyses in order to see if these IP addresses had tried to connect to other ports of our honeypot except for these interactive connections; and the answer is no. Furthermore, the machines that make interactive `ssh` connections on our honeypot do not make any other kind of connections on this honeypot, i.e, no scan or dictionary attack. Further analyses, using the data collected from the low-interaction honeypots deployed in the CADHo project, revealed that none of the 35 IP addresses have ever been observed on any of our platforms deployed in the word. This is interesting because it shows that these machines are totally dedicated to this kind of attack (they only targeted our high-interaction honeypot and only when they knew at least one login and password on this machine).

We can conclude for these analyses that we face two groups of attacking machines. The first group is composed of machines that are specifically in charge of making dictionary attacks. Then the results of these dictionary attacks are published somewhere. Then, another group of machines, which has no intersection with the first group, comes to exploit the weak passwords discovered by the first group. This second group of machines is, as far as we can see, clearly geographically identified and commands are executed by a human being. A similar two steps process was already observed in the CADHo project when analyzing the data collected from the low-interaction honeypots (see [9] for more details).

### 3.3. Behavior of attackers

This section is dedicated to the analysis of the behavior of the intruders. We first characterize the intruders, i.e. we try to know if they are humans or programs. Then, we present in more details the various actions they have carried out on the honeypot. Finally, we try to figure out what their skill level seems to be.

We concentrate the analyses on the last three months of our experiment. During this period, some intruders have visited our honeypot only once, others have visited it several times, for a total of 38 `ssh` intrusions. These intrusions were initiated from 16 IP addresses and 7 accounts were used. Table 3 presents the number of intrusions per account, IP addresses and passwords used for these intrusions. It is of course difficult to be sure that all the intrusions for a same account are initiated by the same person. Nevertheless, in our case, we noted that:
- most of the time, after his first login, the attacker changes the weak password into a strong which, from there on, remains unchanged.
- when two different IP addresses access the same account (with the same password), they are very close and belong to the same country or company.

These two remarks lead us to believe that there is in general only one person associated to the intrusions for a particular account.

| Account | Number of intrusions | Number of passwords | Number of IP addresses |
|---------|---------------------|---------------------|------------------------|
| UA2     | 1                   | 1                   | 1                      |
| UA4     | 13                  | 2                   | 2                      |
| UA5     | 1                   | 1                   | 1                      |
| UA8     | 1                   | 1                   | 1                      |
| UA10    | 9                   | 2                   | 2                      |
| UA13    | 6                   | 1                   | 5                      |
| UA16    | 5                   | 1                   | 3                      |
| UA17    | 2                   | 1                   | 1                      |

**Table 3: Number of intrusions per account**

**3.3.1. Type of the attackers: humans or programs.** Before analyzing what intruders do when connected, we can try to identify who they are. They can be of two different natures. Either they are humans, or they are programs which reproduce simple behaviors. For all intrusions but 12, intruders have made mistakes when typing commands. Mistakes are identified when the intruder uses the backspace to erase a previously entered character. So, it is very likely that such activities are carried out by a human, rather than programs.

When an intruder did not make any mistake, we analyzed how the data were transmitted from the attacker machine to the honeypot. We can note that, for `ssh` communications, data transmission between the client and the server is asynchronous. Most of the time, the `ssh` client implementation uses the function `select()` to get user input. So, when the user presses a key, this function ends and the program sends the corresponding value to the server. In the case of a copy and a paste into the terminal running the client, the `select()` function also ends, but the program sends all the values contained in the buffer used for the paste into the server. We can assume that, when `tty_read()` returns more than one character, these values have been sent after a copy and a paste. If all the activities during a connection are due to a copy and a paste, we can strongly assume that it is due to an automatic script. Otherwise, this is quite likely a human being who uses shortcuts from time to time (such as CTRL-V to paste commands into its `ssh` session). For 7 out of the last 12 activities without mistakes, intruders have entered several commands on a character-by-character basis. This, once again, seems to indicate that a human being is entering the commands. For the 5 others, their activities are not significant enough to conclude: they have only launched a single command, like w, which is not long enough to highlight a copy and a paste.

**3.3.2. Attacker activities.** The first significant remark is that all of the intruders change the password of the hacked account. The second remark is that most of them start by downloading some files. In all cases, but one, the attackers tried to download some malware to the compromised machines. In a single case, the attacker has first tried to download an innocuous, yet large, file to the machine (the binary for a driver coming from a known web site). This is probably a simple way to assess the connectivity quality of the compromised host.

The command used by the intruders to download the software is `wget`. To be more precise, 21 intrusions upon 38 include the `wget` command. These 21 intrusions concern all the hacked accounts. As mentioned in section 2, outgoing `http` connections are forbidden by the firewall. Nevertheless, the intruders still have the possibility to download files through the `ssh` connection using `sftp` command (instead of `wget`). Surprisingly, we noted that only 30% of the intruders did use this `ssh` connection. 70% of the attackers were unable to download their malware due to the absence of `http` connectivity! Three explanations can be envisaged at this stage. First, they follow some simplistic cookbook and do not even known the other methods at their disposal to upload a file. Second, the machines where the malware resides do not support `sftp`. Third, the lack of `http` connectivity made the attacker suspicious and he decided to leave our system. Surprisingly, the first explanation seems to be the right one in our case as we noticed that the attackers leave after an unsuccessful `wget` and come back a few hours or days later, trying the same command again as if they were hoping it to work at that time. Some of them have been seen trying this several times. It can be concluded that: i) they are apparently unable to understand why the command fails, ii) they are not afraid to come back to the machine despite the lack of `http` connectivity, iii) applying such brute force attack reveals that they are not aware of any other method to upload the file.

Once the attackers manage to download their malware using `sftp`, they try to install it (by decompressing or extracting files for example). 75% of the intrusions that installed software did not install it on the hacked account but rather on standard directories such as `/tmp`, `/var/tmp` or `/dev/shm` (which are directories with write access for everybody). This makes the hacker activity more difficult to identify because these directories are regularly used by the operating system itself and shared by all the users.

Additionally, we have identified four main activities of the intruders. The first one is launching `ssh` scans on other networks but these scans have never tested local machines. Their idea is to use the targeted machine to scan other networks, so that it is more difficult for the administrator of the targeted network to localize them. The program used by most intruders, which is easy to find on the Internet, is `pscan.c`.

The second type of activity consists in launching `irc` clients, e.g., `emech` [13] and `psyBNC`. Names of binary files have regularly been changed by intruders, probably in order to hide them. For example, the binary files of `emech` have been changed to `crond` or `inetd`, which are well known Unix binary file names and processes.

The third type of activity is trying to become root. Surprisingly, such attempts have been observed for 3 intrusions only. Two rootkits were used. The first one exploits two vulnerabilities: a vulnerability which concerns the Linux kernel memory management code of the `mremap` system call [14] and a vulnerability which concerns the internal kernel function used to manage process's memory heap [15]. This exploit could not succeed because the kernel version of our honeypot does not correspond to the version of the exploit. The intruder should have realized this because he checked the version of the kernel of the honeypot (`uname -a`). However, he launched this rootkit anyway and failed. The other rootkit used by intruders exploits a vulnerability in the program `ld`. Thanks to this exploit, three intruders became `root` but the buffer overflow succeeded only partially. Even if they apparently became `root`, they could not launch all desired programs (removing files for example caused access control errors).

The last activity observed in the honeypot is related to phishing activities. It is difficult to make precise conclusions because only one intruder has attempted to launch such an attack. He downloaded a forged email and tried to send it through the local `smtp` agent. But, as far as we could understand, it looked like a preliminary step of the attack because the list of recipient emails was very short. It seems that is was just a preliminary test before the real deployment of the attack.

**3.3.3. Attackers skill.** Intruders can roughly speaking be classified into two main categories. The most important one is relative to *script kiddies*. They are inexperienced *hackers* who use programs found on the Internet without really understanding how they work. The next category represents intruders who are more dangerous. They are named "black hat". They can make serious damage on systems because they are expert in security and they know how to exploit vulnerabilities on various systems.

As already presented in §3.3.2. (use of `wget` and `sftp`), we have observed that intruders are not as clever as expected. For example, for two hacked accounts, the intruders don't seem to really understand the Unix file access rights (it's very obvious for example when they try to erase some files whereas they don't have the required privileges). For these two same accounts, the intruders also try to kill the processes of other users. Many intruders do not try to delete the file containing the history of their commands or do not try to deactivate this history function (this file depends on the login shell used, it is `.bash_history` for example for the `bash`). Among the 38 intrusions, only 14 were cleaned by the intruders (11 have deactivated the history function and 3 have deleted the `.bash_history` file). This means that 24 intrusions left behind them a perfectly readable summary of their activity within the honeypot.

The IP address of the honeypot is private and we have started another honeypot on this network. This second honeypot is not directly accessible from the outside, it is only accessible from the first honeypot. We have modified the `/etc/motd` file of the first honeypot (which is automatically printed on the screen during the login process) and added the following message: "`In order to use the software XXX, please connect to A.B.C.D`". In spite of this message, only one intruder has tried to connect to the second honeypot. We could expect that an experienced hacker will try to use this information. In a more general way, we have very seldom seen an intruder looking for other active machines on the same network.

One important thing to note is relative to fingerprinting activity. No intruder has tried to check the presence of VMware software. For three hacked accounts, the intruders have read the contents of the file `/proc/cpuinfo` but that's all. None of the methods discussed on Internet was tested to identify the presence of VMware software [16,17]. This probably means that the intruders are not experienced hackers.

## 4. Conclusion

In this paper, we have presented the results of an experiment carried out over a period of 6 months during which we have observed the various steps that lead an attacker to successfully break into a vulnerable machine and his behavior once he has managed to take control over the machine.

The findings are somehow consistent with the informal know how shared by security experts. The contributions of the paper reside in performing an experiment and rigorous analyses that confirm some of these informal assumptions. Also, the precise analysis of the observed attacks reveals several interesting facts. First of all, the complementarity between high and low interaction honeypots is highlighted as some explanations can be found by combining information coming from both set ups. Second, it appears that most of the observed attacks against port 22 were only partially automated and carried out by script kiddies. This is very different from what can be observed against other ports, such as 445, 139 and others, where worms have been designed to completely carry out the tasks required for the infection and propagation. Last, honeypot fingerprinting does not seem to be a high priority for attackers as none of them has tried the known techniques to check if they were under observation. It is also worth mentioning a couple of important missing observations. First, we did not observe scanners detecting the presence of the open ssh port and providing this information to other machines in charge of running the dictionary attack. This is different from previous observations reported in [9]. Second, as most of the attacks follow very simple and repetitive patterns, we did not observe anything that could be used to derive sophisticated scenarios of attacks that could be analyzed by intrusion detection correlation engine. Of course, at this stage it is too early to derive definite conclusions from this observation.

Therefore, it would be interesting to keep doing this experiment over a longer period of time to see if things do change, for instance if a more efficient automation takes place. We would have to solve the problem of weak passwords being replaced by strong ones though, in order to see more people succeeding in breaking into the system. Also, it would be worth running the same experiment by opening another vulnerability into the system and verifying if the identified steps remain the same, if the types of attackers are similar. Could it be, at the contrary, that some ports are preferably chosen by script kiddies while others are reserved to some more elite attackers? This is something that we are in the process of assessing.

**Acknowledgement.** This work has been partially supported by: 1) CADHo, a research action funded by the French ACI "Securité & Informatique" (www.cadho.org), 2) the CRUTIAL IST-027513 project (crutial.cesiricerca.it), and 3) the ReSIST IST- 026764 project (www.resist-noe.org).